\documentclass[aip,graphicx,amsmath,amssymb,onecolumn]{revtex4-1}

\usepackage{graphicx}
\usepackage{color}
\usepackage{bm}
\usepackage{stackrel}
\usepackage{setspace}

\newcommand{\mb}[1]{\textcolor{blue}{#1}}

\begin{document}
\title{\Large{\sc Response to Halatek and Frey:}\\Effective two-dimensional model does account for geometry sensing 
by self-organized proteins patterns}
\author{Mike Bonny}
\affiliation{Theoretical Physics, Saarland University, Saarbr\"ucken, Germany}
\author{Jakob Schweizer}
\affiliation{Biophysics, BIOTEC, Dresden University of Technology, Dresden, Germany}
\author{Martin Loose}
\affiliation{Biophysics, BIOTEC, Dresden University of Technology, Dresden, Germany}
\affiliation{Department of Systems Biology, Harvard Medical School, Boston, USA}
\author{Ingolf M\"onch}
\affiliation{Institute for Integrative Nanosciences, IFW Dresden, Dresden, Germany}
\author{Petra Schwille}
\affiliation{Biophysics, BIOTEC, Dresden University of Technology, Dresden, Germany}
\affiliation{Max-Planck-Institute for Biochemistry, 82152 Martinsried, Germany}
\author{Karsten Kruse}
\affiliation{Theoretical Physics, Saarland University, Saarbr\"ucken, Germany}

\begin{abstract}
\begin{singlespace}
The Min proteins from Escherichia coli can self-organize into traveling waves on supported lipid bilayers.
In Ref.~[\onlinecite{schw12}] we showed that these waves are guided along the boundaries of membrane patches. We 
introduced an effective two-dimensional model reproducing the observed patterns. In their text~\cite{hala12}, 
Jacob Halatek and Erwin Frey contest the ability of our effective two-dimensional model to describe the dynamics of Min 
proteins on patterned supported lipid bilayers.  
We thank Halatek and Frey for their interest in our work and for again highlighting the importance of dimensionality and
geometry for pattern
formation by the Min proteins. Here we reply in detail to the objections by Halatek and Frey and show that (1) our effective two-dimensional model 
 reproduces the observed patterns on isolated patches and that (2) a three-dimensional version of our model produces similar patterns
on square patches.
\end{singlespace}
\end{abstract}
\maketitle

\section{Brief response to Halatek and Frey}

Here, we summarize the essential points of our response to Halatek and Frey. A detailed response can be found below.\\[1mm]

Halatek and Frey~\cite{hala12} contest our claim that a two-dimensional effective model of Min-protein dynamics accounts for 
geometry sensing by self-organized patterns~\cite{schw12}. Our model notably includes transient binding of MinE to the membrane, a process 
that to our knowledge was first suggested by Meacci and Kruse~\cite{meac05} and then studied in context of MinE-ring formation by Derr et 
al.~\cite{derr09}. The impact of MinE membrane binding on Min-protein patterns in a cellular geometry was first studied subsequently by 
Arjunan and Tomita~\cite{arju10}.

Hsieh et al. initially suggested direct membrane binding of MinE, which has been attributed to positively charged residues located at positions 10-12 of 
MinE~\cite{hsie10}. In Loose et al, we mutated these residues and found that the corresponding mutant MinE 
(C1) still supports dynamic instability of the Min system in vitro, but fails to displace MinC from MinD~\cite{loos11}. In a subsequent structural study 
by Park et al.~\cite{park11} a nascent amphipathic helix at the extreme N-terminus of MinE was identified, which allows MinE to directly interact 
with the membrane. In contrast to the study by Hsieh et al, the residues involved in membrane binding were found to be located at positions 3-8. 
Park et al. suggested that residues 10-12 might still contribute to membrane-binding, but are more important for the interaction with MinD since 
mutations here impair MinE's ability to displace MinC from MinD. In light of this more recent data, we conclude that the MinE C1 mutant we 
tested still has some affinity to the membrane. So far, we have not been able to study the effect of mutations of residues 3-8 in our in vitro assay, 
however it was found they lead to erratic oscillations in vivo (J. Lutkenhaus personal communication). We therefore conclude that MinE 
membrane binding is required for the formation of regular Min protein patterns and consequently geometry sensing.

In contrast to the allegations by Halatek and Frey the mechanism studied in Ref.~\onlinecite{schw12} does account for the observed 
MinDE protein patterns. This holds for a two- dimensional as well as for a three-dimensional implementation, see below. The parameters 
presented in Ref.~\onlinecite{schw12} were unfortunate in that the corresponding solutions depended on the size of the gold layer 
implemented in the simulations. As the data presented below show, however, this is not a short-coming of either the mechanism nor the model. 
Furthermore, the purpose of our model was to show that accounting for transient MinE membrane binding allows us to reproduce the effect of 
geometry sensing observed in experiments. Parameters were not optimized to yield quantitative agreement with experiments or to reproduce 
the exact form of the density profiles. More importantly and in contrast to all other published models, however, the mechanism we studied is 
able to qualitatively reproduce all hitherto observed Min-protein patterns.

The two-dimensional model studied in Ref.~\onlinecite{schw12} is an effective description of the Min-protein dynamics in three dimensions. 
%However, s
Solving the corresponding model equations in three dimensions can also produce patterns similar to those observed in experiment. 
Although a formal mapping between the three-dimensional and the two-dimensional model is still lacking, we find that there is
a correspondence between the patterns generated by both models. This feature highlights the robustness of the mechanism that
we propose to explain the essential features of the Min-protein dynamics observed in vivo and in vitro.
%There are parameter values, such that close to the top and the side boundaries of the simulation domain, the distributions of MinD and MinE 
%in the buffer are essentially homogeneous. In these cases, an additional enlargement of the system would not affect the pattern on the membrane 
%patch in an essential way.

In conclusion, while our choice of parameters might have been unfortunate in some respects, we maintain our claim that transient membrane 
binding of MinE is able to reproduce the experimentally observed geometry sensing. 

\section{Detailed response to Halatek and Frey}

\subsection{Introduction}

In our work~[\onlinecite{schw12}], we presented experimental data showing that spontaneous surface waves of Min 
proteins on supported lipid bilayers can be guided by lateral confinement. In these
experiments binding of the Min proteins to the surface was restricted to membrane patches of different shapes and sizes by 
covering the rest of the surface with a passivating gold layer. The patches were either arranged in arrays or isolated, separated
by about 50 $\mu$m from each other. We analyzed the Min-protein waves on the membrane patches by 
using an effective two-dimensional description of the protein dynamics on the surface. In this effective model, we accounted 
for persistent binding of MinE to the membrane. While this property had already been proposed to be important for the formation 
of the so-called E-ring [\onlinecite{derr09}]-[\onlinecite{hu02}],
an accumulation of MinE at the rim of the polar MinD zone in vivo, experimental evidence for this assumption has only been available recently.

Our model describes the evolution of the densities $c_D$ and $c_E$ of MinD and MinE not bound to the membrane and 
of the densities $c_d$, $c_{de}$, and $c_e$ of membrane-bound MinD, MinDE complex, and MinE, respectively, using a
meanfield approach. In our deterministic, effective two-dimensional description, the densities are all surface densities. The 
densities $c_d$, $c_{de}$, and $c_e$ are restricted to the regions of the membrane patches. The complete system, however, is
larger and comprises also regions corresponding to the gold layer. The latter regions are only accessible to MinD and MinE in
solution. Explicitly, the equations read
\begin{align}
\label{eq:cDdot}
\partial_t c_D & = D_D\Delta c_D-c_D(\omega_D+\omega_{dD}c_d)(c_\mathrm{max}-c_d-c_{de})/c_\mathrm{max}\nonumber\\
&\quad+(\omega_{de,m}
+\omega_{de,c})c_{de}\\
\partial_t c_E & = D_E\Delta c_E-\omega_Ec_Ec_d+\omega_{de,c}c_{de}+\omega_ec_e\\
\label{eq:cddot}
\partial_t c_d & = D_d\Delta c_d + c_D(\omega_D+\omega_{dD}c_d)(c_\mathrm{max}-c_d-c_{de})/c_\mathrm{max}\nonumber\\
&\quad -\omega_Ec_Ec_d-\omega_{ed}c_ec_d\\
\partial_t c_{de} & = D_{de}\Delta c_{de}+\omega_Ec_Ec_d+\omega_{ed}c_ec_d-(\omega_{de,m}
+\omega_{de,c})c_{de}\\
\label{eq:cedot}
\partial_t c_e & = D_e\Delta c_e + \omega_{de,m}c_{de}-\omega_{ed}c_ec_d-\omega_ec_e\quad.
\end{align}
For the diffusion currents of MinD and MinE not attached to the membrane, we apply periodic boundary conditions at the system's 
boundaries. The membrane patches do not affect diffusion of the unbound proteins. In addition, we apply zero flux 
conditions on the currents for the membrane-bound densities perpendicular to the patch boundaries. The various constants 
$D_\alpha$ denote the diffusions constants for species $\alpha=D$, $E$, $d$, $de$, and $e$. The values of $\omega_D$ 
and $\omega_{dD}$ parameterize binding of MinD to the membrane, $\omega_E$ binding of MinE. Dissociation of MinDE 
complexes is described by $\omega_{de,m}$ and $\omega_{de,c}$ giving unbinding of MinD with MinE staying on the
membrane or not, respectively. Finally, $c_\mathrm{max}$ is the maximal protein density on the surface. 

Let us point out that this description does not account for possible intermolecular interactions between membrane-bound molecules.
There is some experimental evidence for such interactions~\cite{hu02,shih05,loos11} and they have been shown to be able to 
trigger a dynamic instability that leads to pattern formation of Min proteins~\cite{krus02,meac05}. However, these interactions
have not been characterized experimentally and their molecular nature is presently unknown. We thus refrain at this point from making any
ad hoc assumptions. As we will see below, these suggested interactions are not necessary to comprehensibly describe the 
influence of the geometry on the Min-protein patterns nor for giving a comprehensive description of all Min-protein patterns observed in
vivo up to date~\cite{bonn13}. They are presumably important, though, to get a quantitative agreement
of the wave form and the residence times of the Min proteins on the membrane as we will see below.

The solutions to these equations we presented in Ref.~[\onlinecite{schw12}] were obtained for parameter values that present
certain shortcomings. However, as we show here, these shortcomings are indeed linked to the parameter values and not to the
model per se. Since a comprehensive numerical study of parameter space is prohibited by
limited computer power. Further quantitative experiments are needed to provide constraints on parameters. For the time
being it seems appropriate to focus on certain features of the Min-protein patterns. 
We will come back to this point in the discussion at the end of this article.

We will now give a point-to-point response to the issues raised by Halatek and Frey. For convenience of the reader, the points 
of Halatek and Frey are copied in this text.

\subsection{Response to the points made by Halatek and Frey}

\noindent\textit{\mb{We investigated the simulation files provided by the authors and found that the model neither accounts for actual MinE membrane interactions nor for any observed MinDE protein patterns. It does not reproduce any of the computational data presented in the article [\onlinecite{schw12}].}}\\
In contrast to the allegations by Halatek and Frey, the computational data were obtained from the dynamic equations given above and in the article. Our model includes the densities $c_{de}$ and $c_e$, which describe MinE bound to the membrane either in a complex with MinD or alone. The figures presented in Ref.~[\onlinecite{schw12}] show patterns similar to the ones observed in the corresponding experiments. In Ref.~[\onlinecite{bonn13}], we furthermore show, that our model reproduces all experimental Min-protein patterns reported so far.\\[1mm]

\noindent\textit{\mb{For the published parameters, pattern formation is restricted to very small cytosol/membrane ratios.}}\\
It is not clear to us, which "cytosol/membrane ratios" Halatek and Frey have in mind. In any case, in Ref.~\onlinecite{schw12}, 
we make no statement about the parameter dependence of the in vitro patterns.\\[1mm]

\noindent\textit{\mb{Cytosolic volume is not accounted for and total densities indicate an effective bulk height below $6\mu m$.}}\\
Indeed, in Ref.~\onlinecite{schw12} we considered an effective 2d model as has been done in several studies before our work. The relation 
between the full
three-dimensional model and the effective 2d model is not obvious and remains to be worked out. For the time being we can say,
though, that both models produce similar patterns for the membrane-bound protein densities. Below, we give results 
that show that the mechanism studied by our computational analysis in 2d and 3d yields patterns similar to the experimental patterns.
In our 3d calculations we used a bulk height of 90$\mu$m and we observed that the cytosolic concentrations were essentially homogenous
above a height of about 15$\mu$m. Furthermore, as shown in Fig.~S1 in the appendix of Ref.~[\onlinecite{schw12}], the protein patterns 
are indeed confined to a narrow region above the membrane above which the corresponding fluorescence intensitiy appears to be 
homogenous.\\[1mm]

\noindent\textit{\mb{We find that scaling the cytosolic dynamics by a small factor $\mathcal O(1)$ or increasing gold layer size eliminates the instability. Hence, the model configuration deviates from the experiment by orders of magnitude. In striking contradiction to the accompanying experiments and to the claim in the article, bulk volume has a severe effect on the computational model.}}\\
We are not sure, what Halatek and Frey have in mind here. In our two-dimensional computational model we cannot study the effect of the bulk volume on the patterns formed, because we do not have an exact mapping of the 3d model to 2d. We show below that in a full three-dimensional model
of the same mechanism as used for the two-dimensional model in Ref.~[\onlinecite{schw12}], the bulk volume does not
affect the membrane patterns as long as it is higher than 15$\mu$m.\\[1mm]

\begin{figure}
\includegraphics[width=0.7\textwidth]{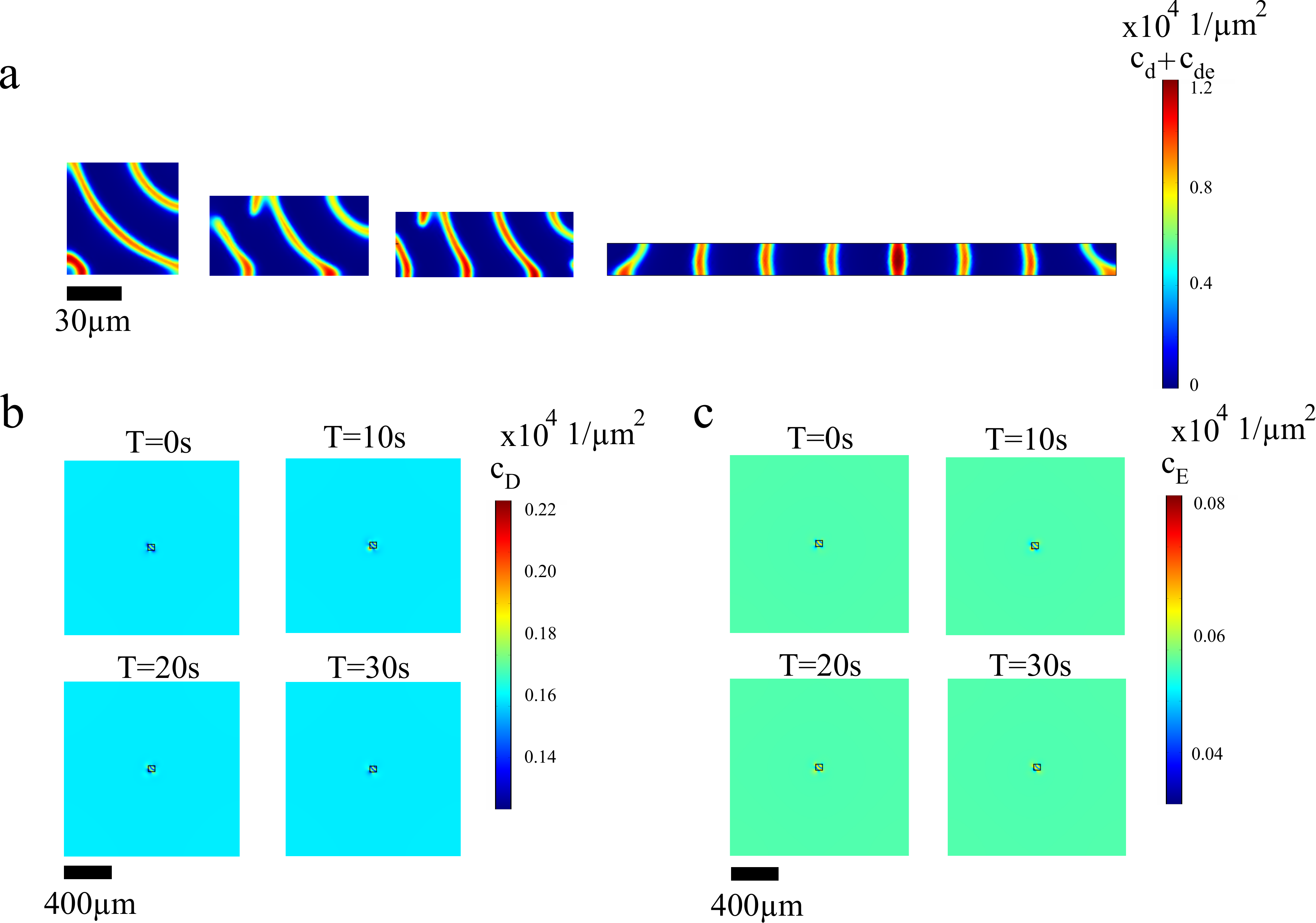}
\caption{\label{fig:AspectRatios2D} Solutions of the 2d model equations on rectangular membrane patches for 
the parameters given in Table~\ref{tab:parametersInVitro}. a) Snapshots of the distribution of membrane-bound MinD on rectangular 
membrane patches with different aspect ratios. In the rightmost case, planar waves emanating from the two opposite sides
meet in the patch center. In the other cases, waves emanate from the upper right corner. b) Distribution of MinD in the buffer for the
square membrane patch shown in (a). c) Distribution of MinE in the buffer for the square membrane patch shown in (a). In (b) and (c) 
the whole simulated area is shown for the patch with aspect ratio 1:1. Heterogeneities in the concentrations are limited to 
the region of the membrane patch. Similar behavior is observed for the other patch geometries.}
\end{figure}

\begin{figure}
\includegraphics[width=0.7\textwidth]{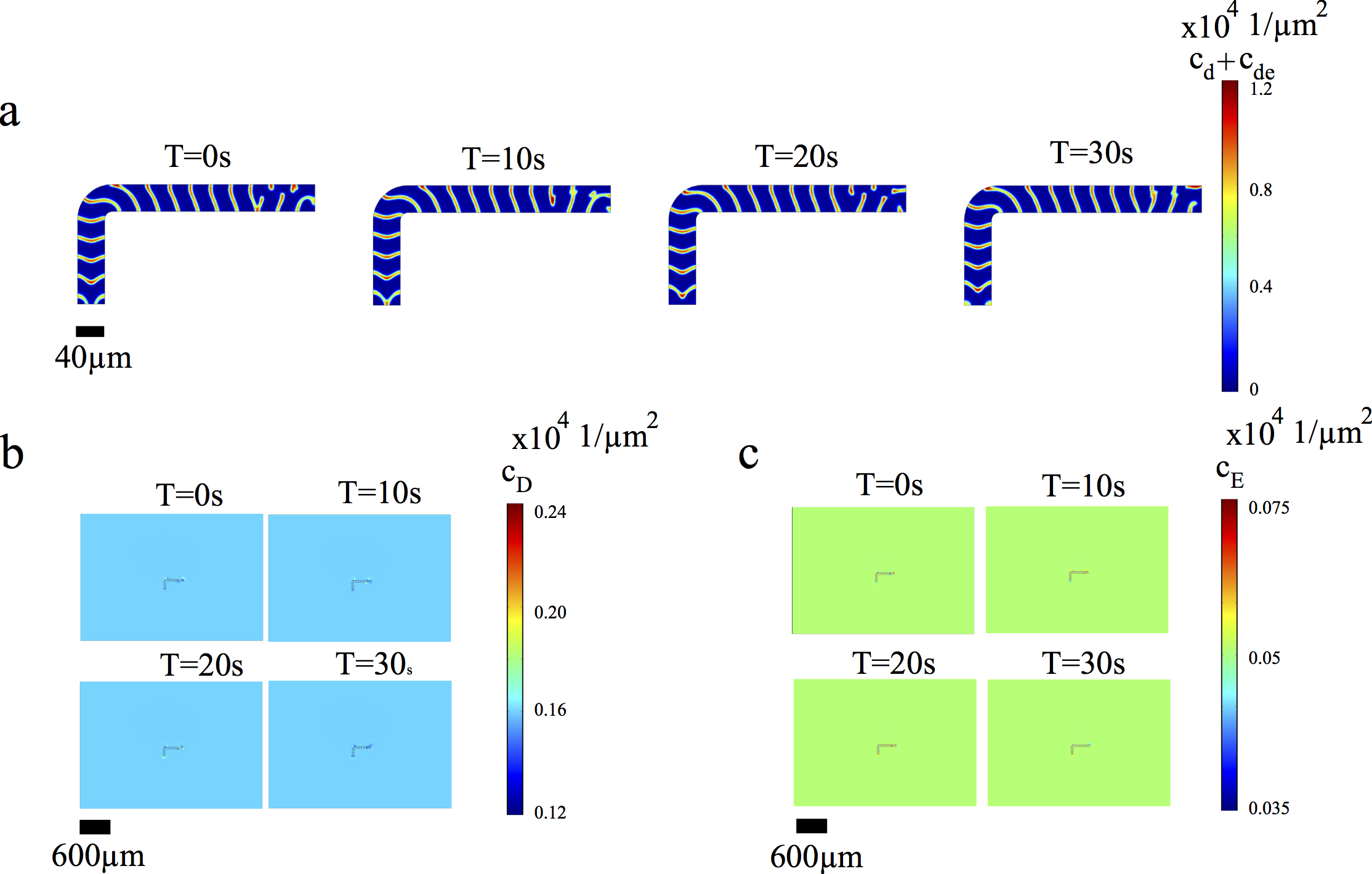}
\caption{\label{fig:RightAngle} Solution of the 2d model equations in presence of an L-shaped membrane patch 
for the parameters given in Table~\ref{tab:parametersInVitro}. a) Snapshots of the distribution of membrane-bound MinD on the
membrane patch. Waves are guided by the patch an move from left to right. b, c) Corresponding distributions of MinD and MinE
in buffer. Again heterogeneities in the concentrations are limited to the region of the membrane patch.}
\end{figure}

\begin{figure}
\includegraphics[width=0.8\textwidth]{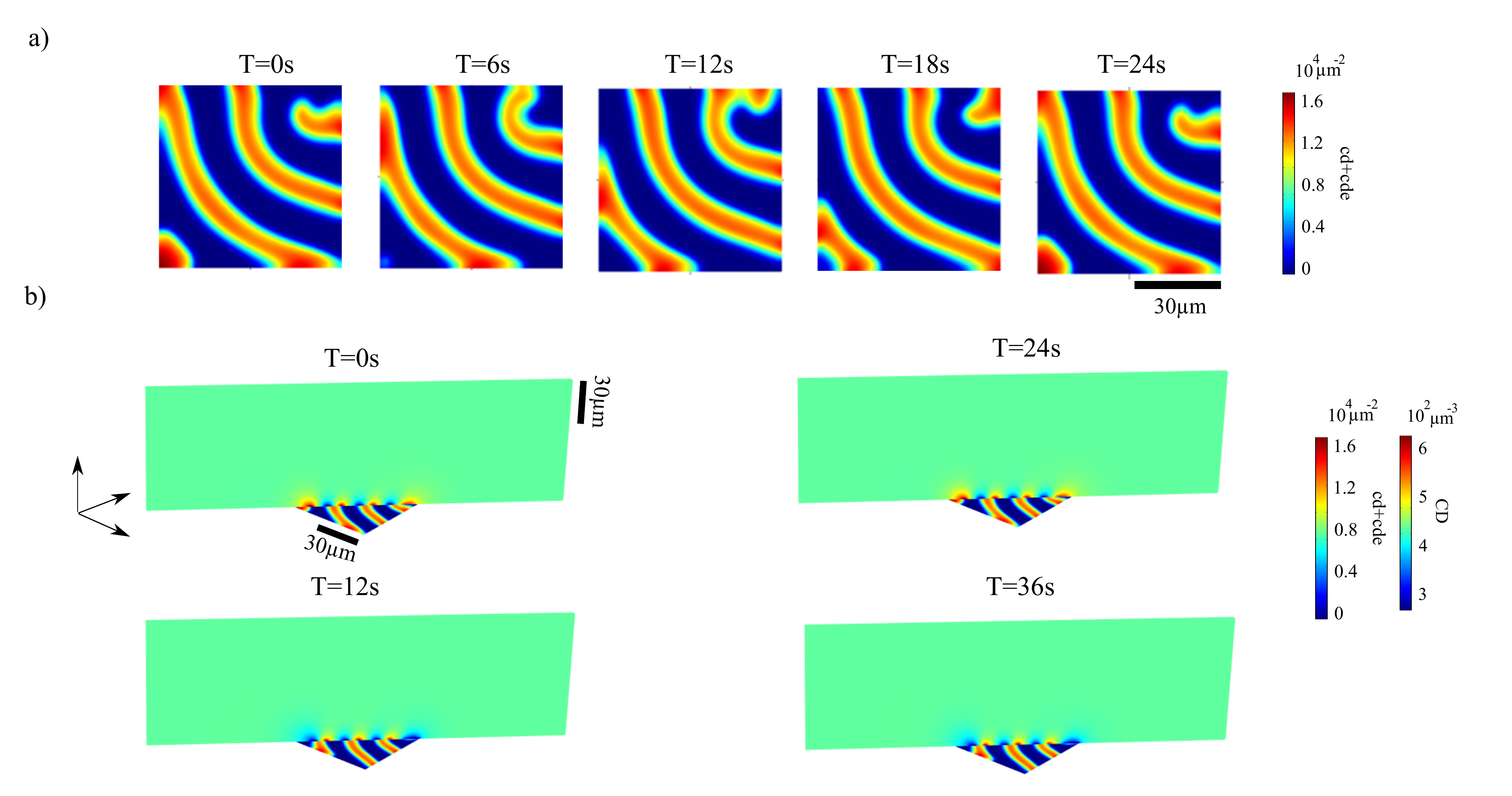}
\caption{\label{fig:SquarePatch3D}Solution of the 3d model equations in presence of a quadratic membrane patch. a) Distribution of 
membrane-bound MinD on the patch. b) Buffer concentration of MinD along a representative slice through the system. Compare to Fig.~S1 
in the appendix of Ref.~[\onlinecite{schw12}]. Parameters are given in Table~\ref{tab:parametersInVitro3d}.}
\end{figure}

\begin{figure}
\includegraphics[width=0.8\textwidth]{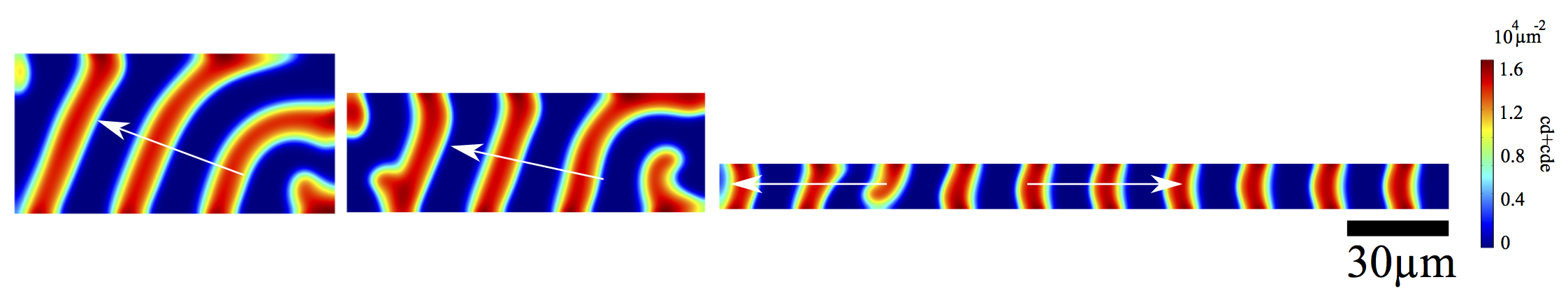}
\caption{\label{fig:Patch3DAspectRatio}Solution of the 3d model equations on rectangular membrane patches with different aspect ratios.
The cytosolic concentrations are not shown. Parameters are given in Table~\ref{tab:parametersInVitro3d}.}
\end{figure}

\begin{figure}
\includegraphics[width=0.8\textwidth]{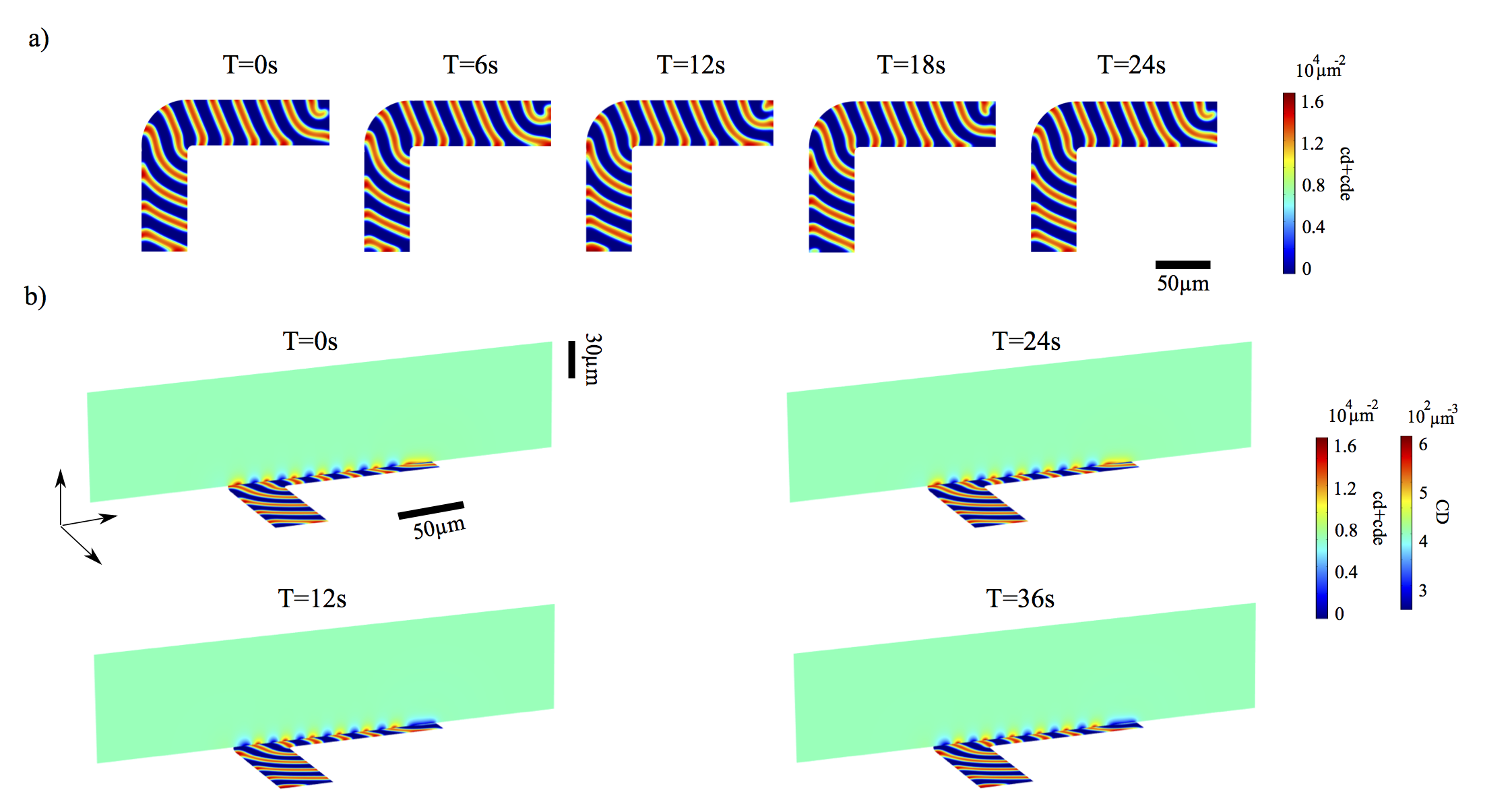}
\caption{\label{fig:AspectRatio3D}Solution of the 3d model equations in presence of a L-shaped membrane patch. a) Distribution of 
membrane-bound MinD on a L-shaped patch. b) Buffer concentration of MinD along  a representative slice through the system. Compare to 
Fig.~S1 in the appendix of Ref.~[\onlinecite{schw12}].
Parameters are given in Table~\ref{tab:parametersInVitro3d}.}
\end{figure}

\noindent \textit{\mb{The authors compensated for this system size dependence by adjusting intrinsic system parameters (MinE/MinD ratio)
without mentioning it. Moreover, the adjusted parameters deviate from the experimental value while the published parameters do not.}}\\
For the parameter sets given below for the calculations in 2d and 3d,
our computational model produces all observed patterns, see Figs.~\ref{fig:AspectRatios2D}-\ref{fig:AspectRatio3D}. In the 3d calculations,
the total MinD and MinE densities correspond to the values used in Ref.~[\onlinecite{loos11}].\\[1mm]

\noindent \textit{\mb{Even with these adjustments the model relies on simulation artifacts to reproduce the experimental data. Alignment 
to the aspect ratio requires periodic boundaries at the gold layer. The alignment angle is controlled by cross-boundary coupling in horizontal 
and vertical directions. The aspect ratio of the patch has a negligible effect on alignment. Without periodic boundaries or for gold layers 
sizes as used in the experiment alignment ceases and waves become disordered blobs. This invalidates the model on a conceptual level.}}\\
As shown in Figures \ref{fig:AspectRatios2D} and \ref{fig:RightAngle} below, our computational model (or rather the molecular processes 
implemented in the model)
can reproduce the experimentally observed patterns on patched membranes in a robust way and independently of the boundary conditions
chosen for the gold layer.
In Figures~\ref{fig:AspectRatios2D} and \ref{fig:RightAngle}, we present numerical solutions that were obtained for patches surrounded
by an inert region of sizes 1.625mm$\times$1.625mm and 2.303mm$\times$3.290mm, using the parameter set given in 
Table~\ref{tab:parametersInVitro} for the 2d model defined above. Solutions were obtained by using 
Comsol Multiphysics 4.1\textsuperscript{\textregistered}. Far away from the patch, the densities $c_D$ and $c_E$ are essentially 
homogenous and the size of the surrounding gold layer can be increased further without affecting the pattern on the patch. 
While we agree that the parameter set used in Ref.~[\onlinecite{schw12}] was unfortunate as the patterns depend for
these values on the size of the gold layer, we conclude 
that the mechanism studied in our work is indeed capable of reproducing the patterns observed experimentally
also if the membrane patches are isolated and thus is valid on a conceptual level.\\
\begin{center}
\begin{table}[h]
\begin{tabular}{|c|c|c|c|c|c|c|c|}
 $D_D$ & $D_E$ & $D_d$ & $D_{e}$ & $D_{de}$ & $c_\mathrm{max}$ & $\omega_D$ \\
$50\frac{\mu\mathrm{m}^2}{s}$ & $50\frac{\mu\mathrm{m}^2}{s}$ & $0.24\frac{\mu\mathrm{m}^2}{s}$ & $0.48\frac{\mu\mathrm{m}^2}{s}$ 
& $0.24\frac{\mu\mathrm{m}^2}{s}$ & $2.0\cdot10^4 \frac{1}{\mu \mathrm{m}^2}$ & 0.045$\frac{1}{s}$ \\ 
 $\omega_{dD}$ & $\omega_E$ & $\omega_{ed}$ & $\omega_{de,c}$ & $\omega_{de,m}$ & $\omega_e$ & $C_{D0}$ & $C_{E0}$  \\
 $9\cdot10^{-4}\frac{\mu \mathrm{m}^2}{s}$ &  $4\cdot10^{-4}\frac{\mu \mathrm{m}^2}{s}$ 
& $2.5\cdot10^{-3}\frac{\mu \mathrm{m}^2}{s}$ & 0.08$\frac{1}{s}$ & 0.8$\frac{1}{s}$ & 0.08$\frac{1}{s}$ & 1631 $\frac{1}{\mu\mathrm{m}^2}$ & 563 $\frac{1}{\mu\mathrm{m}^2}$ 
\end{tabular}
\caption{\label{tab:parametersInVitro}
Parameter values for the 2d model used in Figs.~\ref{fig:AspectRatios2D} and \ref{fig:RightAngle}.}
\end{table}
\end{center}

In order to underline this point even further we will now 
show that the molecular mechanisms accounted for in our computational model can produce  patterns similar to those observed 
experimentally also in a three-dimensional geometry.

Let us first describe the formulation of our model in three spatial dimensions. We consider the volume densities $c_{D}$ 
and $c_E$ for MinD and MinE, respectively, and the surface densities $c_d$, $c_{de}$, and $c_e$ for MinD, 
MinDE complexes, and MinE. 
Note, that the surface densities are defined only on the regions of the surface corresponding to membrane patches, but not in the 
regions corresponding the gold layer. The dynamic equations for our model in three spatial dimensions are then given by
\begin{eqnarray}
\label{eq:dcDdt} \partial_t c_{D} & = & D_D\Delta c_{D}\\ % + \omega_{DT} c_{DD}\\ 
%\label{eq:dcDDdt} \partial_t c_{DD} & = & D_D\Delta c_{DD}- \omega_{DT} c_{DD}\\
\label{eq:dcEdt} \partial_t c_E & = & D_E\Delta c_E\\
\label{eq:dcddt} \partial_t c_d & = & D_d\Delta_\parallel c_d + c_{D}(\omega_D+\omega_{dD}c_d)(c_\mathrm{max}-c_d-c_{de})/c_\mathrm{max}  \nonumber \\ & & -\omega_Ec_Ec_d-\omega_{ed}c_ec_d\\
\label{eq:dcdedt} \partial_t c_{de} & = & D_{de}\Delta_\parallel c_{de}+\omega_Ec_Ec_d+\omega_{ed}c_ec_d \nonumber \\ & &-(\omega_{de,m}+\omega_{de,c})c_{de}\\
\label{eq:dcedt} \partial_t c_e & = & D_e\Delta_\parallel c_e + \omega_{de,m}c_{de}-\omega_{ed}c_ec_d-\omega_ec_e\quad.
\end{eqnarray}
The first two equations are defined in the volume above the surface, the remaining equations only in the regions of the surface 
corresponding to the membrane patches. The parameters have the same meaning as in Ref.~[\onlinecite{schw12}]. 
Let us note immediately, however, that it is not obvious how to relate the respective values of the corresponding parameters in the three- and
two-dimensional models. We do observe, though, that both variants of the model produce qualitatively similar patterns if parameters
are chosen appropriately. Future work has to establish the exact relation between the two variants. 

In addition to the bulk equations, we also have to determine the boundary conditions on the diffusion currents. At the boundaries of 
the membrane patches we use no-flux conditions for the surface densities. For the buffer species $c_D$ and $c_E$, the diffusion 
currents perpendicular to the boundaries vanish at the system's top as well as on the gold layer. In the region of the membrane,
the component of the diffusive fluxes perpendicular to the surface obey:
\begin{align}
-D_D\nabla_\perp c_D &= c_D(\omega_D+\omega_{dD}c_d)(c_\mathrm{max}-c_d-c_{de})/c_\mathrm{max}-(\omega_{de,m}+\omega_{de,c})c_{de}\\
-D_E\nabla_\perp c_E &= \omega_Ec_Ec_d-\omega_e c_e-\omega_{de,c}c_{de}.
%-D_D\nabla_\perp c_{DT} &= c_{DT}(\omega_D+\omega_{dD}c_d)(c_\mathrm{max}-c_d-c_{de})/c_\mathrm{max}\\
%-D_D\nabla_\perp c_{DD} &= -(\omega_{de,m}+\omega_{de,c})c_{de}\\
%-D_E\nabla_\perp c_E &= \omega_Ec_Ec_d-\omega_e c_e-\omega_{de,c}c_{de}.
\label{eq:MinECurrBC2}
\end{align}
Finally, we apply periodic boundary conditions at the system's sides.

We solved the equations numerically using Comsol Multiphysics 4.1\textsuperscript{\textregistered} for the parameter values given 
in Table~\ref{tab:parametersInVitro3d} in a domain of size 200$\mu$m$\times$200$\mu$m$\times$90$\mu$m. The membrane patch
is quadratic with edges 60$\mu$m long. The system produces a wave moving along the diagonal of the patch, see 
Fig.~\ref{fig:SquarePatch3D} and Video S1. The buffer concentrations are homogenous at the systems boundaries with exception of the vicinity of 
the membrane patch indicating that further increasing the simulation box will not affect the pattern on the membrane. Upon increasing 
the aspect ratio of the membrane patch, the waves are more and more directed along the patch's long axis, see 
Fig.~\ref{fig:Patch3DAspectRatio}. The solution
for an L-shaped membrane patch is given in Fig.~\ref{fig:AspectRatio3D} and Video S2. Again densities are homogeneous at the system boundaries
making this solution insensitive to a further increase in the system size. 
\begin{center}
\begin{table}[b]
\begin{tabular}{|c|c|c|c|c|c|c|c|}
 $D_D$ & $D_E$ & $D_d$ & $D_{e}$ & $D_{de}$ & $c_\mathrm{max}$ & $\omega_D$ & $$ \\
$50\frac{\mu\mathrm{m}^2}{s}$ & $50\frac{\mu\mathrm{m}^2}{s}$ & $0.3\frac{\mu\mathrm{m}^2}{s}$ & $1.8\frac{\mu\mathrm{m}^2}{s}$ 
& $0.3\frac{\mu\mathrm{m}^2}{s}$ & $2.75\cdot10^4 \frac{1}{\mu \mathrm{m}^2}$ & 5$\cdot 10^{-4} \frac{\mu\mathrm{m}}{s}$ &  \\ 
 $\omega_{dD}$ & $\omega_E$ & $\omega_{ed}$ & $\omega_{de,c}$ & $\omega_{de,m}$ & $\omega_e$ & $C_{D0}$ & $C_{E0}$  \\
 $3.18\cdot10^{-3}\frac{\mu \mathrm{m}^3}{s}$ &  $1.36\cdot10^{-4}\frac{\mu \mathrm{m}^3}{s}$ 
& $4.9\cdot10^{-3}\frac{\mu \mathrm{m}^2}{s}$ & 0.16$\frac{1}{s}$ & 2.52$\frac{1}{s}$ & 0.5$\frac{1}{s}$ & 484 $\frac{1}{\mu\mathrm{m}^3}$ & 696 $\frac{1}{\mu\mathrm{m}^3}$ 
\end{tabular}
\caption{\label{tab:parametersInVitro3d}
Values of the parameters used for the numerical solutions of the 3d model equations.}
\end{table}
\end{center}
As explained before, we did not attempt to match parameters such that the patterns in the three-dimensional model are exactly the 
same as in the two-dimensional model as many aspects of the systems like the nature of protein-protein interactions on the membrane 
are not yet fully characterized.\\[1mm]

\noindent\textit{\mb{The model is claimed to extend and supersede previous models by incorporating experimental evidence regarding MinE membrane interactions [\onlinecite{park11},\onlinecite{loos11}]. We note that MinE membrane binding was already proposed and analyzed by Arjunan and Tomita.}}\\
Transient membrane-binding by MinE was first suggested in Ref.~[\onlinecite{meac05}] to underlie MinE-ring formation and 
was then studied in this context by Derr et al.~\cite{derr09}. Arjunan and Tomita studied a system, in which MinE could directly bind to another 
membrane-bound MinD after inducing detachment of a membrane-bound MinD~\cite{arju10}. In difference to the model presented in 
Ref.~[\onlinecite{schw12}], though, these authors only considered MinE dimers forming a complex with MinD to stay on the membrane; 
in absence of MinD, MinE could not stay bound to the membrane. The work by Arjunan and Tomita was the first to numerically study a 
stochastic model for Min-protein dynamics in a cellular geometry incorporating this feature. It nicely demonstrated the formation of a 
MinE-ring in this setting.\\[1mm]

\noindent\textit{\mb{Moreover, the model contradicts the experimental references in several aspects. Park et al.~[\onlinecite{park11}] have shown 
that unmasking the anti- MinCD domains in MinE$^{F7E/I24N}$ restores the wild type phenotype without membrane binding. In contrast, 
computational patterns are lost if MinE membrane binding is reduced and cannot be recovered by adjusting MinE recruitment.}}\\
Halatek and Frey falsely claim that experiments by Park {\it et al.} disprove our finding that membrane binding is required or responsible for geometry sensing. In their paper, Park {\it et al.} identified the MinE double-mutant MinE$^{F7E/I24N}$, which does not bind to the membrane {\it in vivo} and is supposedly consecutively activated for MinD binding. While it is true that cells with this mutant show similar colonies on agar plates as with the wildtype protein, Park {\it et al.} make no comment about cell morphology, cell length distributions nor the dynamics of the Min proteins {\it in vivo}. Therefore, ''assuming that the MinE$^{F7E/I24N}$ mutant most likely restores pole-to-pole Min oscillations (hence, geometry sensing) without requiring membrane binding'' as Halatek and Frey do in their letter is unfounded. In fact, this MinE mutant could even fail to induce oscillations but lead to a static Min protein distribution  to regulate cell division as seen in {\it B. subtilis}. To conclude, instead of making assumptions about how proteins might behave, it is important to test these MinE mutants for their ability to initiate protein pattern formation {\it in vivo} as well as {\it in vitro}. 

\noindent\textit{\mb{Hence, the model actually implies that MinE membrane binding is required for pattern formation in the first place and 
not for geometry sensing in particular as the paper claims.}}\\
It would be interesting to test experimentally whether MinE membrane binding is required for pattern formation or not. In any case, we
do not see a contradiction in the two statements (MinE membrane binding is required for pattern formation and MinE membrane binding
is necessary for geometry sensing). Furthermore,
using other models we did not find geometry sensing in the absence of MinE membrane binding, so we maintain our claim that 
MinE membrane binding is necessary for geometry sensing. \\[1mm]

\begin{figure}
\includegraphics[width=0.6\textwidth]{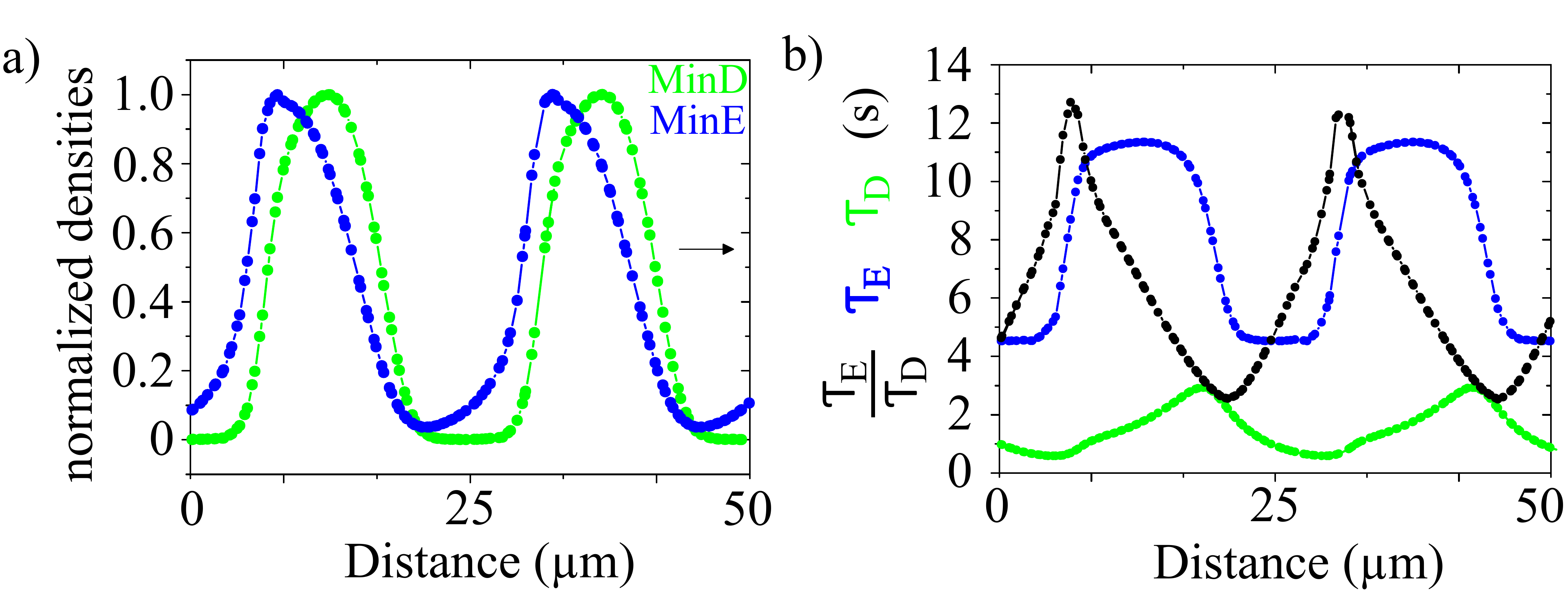}
\caption{\label{fig:residencetime}Properties of traveling waves. a) Profiles of membrane-bound MinD and MinE for the wave 
traveling presented in Fig.~\ref{fig:AspectRatio3D} along the long leg of the L-shaped membrane. b) Spatially resolved residence 
times of MinD and MinE in the membrane and corresponding ratio for the same wave. Parameters are given in 
Table~\ref{tab:parametersInVitro3d}.}
\end{figure}

\noindent\textit{\mb{The ratio of MinE/MinD residence times quantifies the relative strength of MinE membrane binding. It has been 
quantified experimentally by Loose et al.~[\onlinecite{loos11}]. The value in the computational model exceeds the experimental value 
by an order of magnitude. As a consequence MinDE waves contain about ten times more MinE than MinD, in contradiction to 
experiments [\onlinecite{loos11}, \onlinecite{loos08}].}}\\
We estimated the membrane residence times for MinD and MinE in the three-dimensional model for the parameters
given in Tab.~\ref{tab:parametersInVitro3d}. They are given by the inverse of the respective detachment rates. Explicitly, we use
\begin{align}
\langle\tau_{D}\rangle &=(\omega_{E}c_{E}+\omega_{ed}c_{e})^{-1}+(\omega_{de,c}+\omega_{de,m})^{-1}\\
\langle\tau_{E}\rangle&=\frac{2(\omega_{de,m}+\omega_{ed}c_{d}+\omega_{e})}{\omega_{de,c}\omega_{ed}c_{d}+(\omega_{de,c}
+\omega_{de,m})\omega_{e}}.
\end{align}
Because of the non-linear reaction terms, the residence times depend on the protein densities. In Figure~\ref{fig:residencetime},
we present the profile of residence times along a wave traveling along the long bar of the L-shaped membrane presented in 
Fig.~\ref{fig:AspectRatio3D}.  Here, we find an average ratio of the residence times of about 7, which is closer to the experimentally 
reported value than for the parameters reported in the original article, see Fig.~\ref{fig:residencetime}. Due to limited computer power, 
we did not aim at optimizing parameters further.

We would like to emphasize another point: in our meanfield description we did not include interactions between
membrane-bound molecules, which could affect their diffusion constant. Only steric interactions are captured by the maximal 
protein density on the membrane. In contrast to the experimental findings reported in Ref.~[\onlinecite{loos11}],
the diffusion constant of membrane-bound MinD and MinE does thus not change within the wave\footnote{Note also, that in 
Ref.~[\onlinecite{loos11}] the diffusion constant of individual molecules was measured, whereas, in our meanfield description, the 
chemical diffusion constants are used.}. Furthermore, in our description
the residence time of MinD on the membrane decreases as the rear of the wave is approached. This is simply due to an increase
in the concentration of membrane-bound MinE towards the rear of the wave. This is again different from the experimental findings 
reported in Ref.~[\onlinecite{loos11}], where the residence time of MinD increased towards the rear of the wave. As mentioned in 
that work, this observation suggests that intermolecular interactions become more likely at the rear of the wave. These interactions 
could then stabilize the membrane-bound state of MinD. \\[1mm]

\noindent\textit{\mb{In particular, we note that the computational data in Figure 5C cannot be reproduced. We find a 16-fold increased 
MinE/MinD ratio, which represents a 23-fold deviation from the cited experiments [\onlinecite{schw12}].}}\\
Unfortunately, we do not know the results of Halatek and Frey's simulations nor the parameters they used to comment meaningfully
on the discrepancy they
mention here. Concerning the propagation velocity of the waves, which is the quantity discussed here, the relevant concentrations of MinD and
MinE are those at the trailing edge of the wave. In fact, their ratio determines the rate at which MinE removes MinD from the membrane
and thus the propagation velocity. The ratios in Fig.~5C of Ref.~[\onlinecite{schw12}] were read of the density profiles at the position,
where the MinE concentration has reached 85\% of its maximum value. Furthermore, we note that only MinE was labeled in the 
corresponding experiment, such that it remains obscure to us, how Halatek and Frey obtained the ratio of membrane-bound MinE
to membrane-bound MinD from the published experimental data.

\section{Conclusion}

In summary, we maintain that the mechanism encoded in the dynamic equations given above is able to reproduce the 
patterns observed on membrane patches reported in Ref.~[\onlinecite{schw12}]. The mechanism is also able to reproduce
the in vivo patterns as we describe in Ref.~[\onlinecite{bonn13}]. It furthermore predicts the existence of traveling waves in 
sufficiently long cells, which we found also experimentally. This supports the important role of the persistent interaction of MinE 
with the membrane as we concluded in Ref.~[\onlinecite{schw12}]. As supported by the findings by Park et al. this persistent 
interaction of MinE is most likely due to direct, though transient membrane binding of MinE via an N-terminal amphipathic helix. 
A detailed analysis of the model and notably of the relation between the
effective description in two spatial dimensions and the model in three spatial dimensions remains to be done. Furthermore,
more experimental input will be needed to constrain parameter values, notably those describing interactions between 
membrane-bound proteins, 
such that a meaningful quantitative comparison 
between the model calculations and experimental observations will be possible. In this context, one should note, in 
particular, that -- to our knowledge -- none of the models presented prior to the one introduced in Ref.~[\onlinecite{schw12}] 
had been reported to qualitatively reproduce all observed Min-protein patterns \textit{in vivo} as well as in reconstitution 
experiments, whereas the model of Ref.~[\onlinecite{schw12}] does~\cite{bonn13}. We are thus convinced that our model is a 
good starting point for attempting also a quantitative description. 
Obviously, as we learn more about the system, further modifications of the equations might be necessary and parameter
values will be optimized. As stated above, notably, intermolecular interactions could account for some features of the dynamics
of membrane-bound Min proteins. Independently of possibly necessary changes of the model equations, limited computational 
resources and/or the lack of sufficient experimental constraints currently prohibit the identification of a parameter set that would 
reproduce at the same time, wave forms, speeds, lengths, and protein residence times. 

The aim of the theoretical part of Ref.~[\onlinecite{schw12}] was to show that the mechanism represented
by the dynamic equations given above and for which all incorporated processes have experimental support can account  
for the observed guidance of Min-protein surface waves. From the analysis presented above we conclude that these assertions 
remain valid.

\end{document}